\documentclass[twocolumn,tighten]{aastex62}

\usepackage{amsmath}
\usepackage{xspace}
\usepackage{multirow}
\usepackage{fancyapj}

\newcommand{\rxte}{\textit{RXTE}\xspace}

\newcommand{\nicer}{\textit{NICER}\xspace}

\newcommand{\osix}{4U~0614+09\xspace}

\begin{document}

\title{NICER detects a soft X-ray kilohertz Quasi-periodic Oscillation in 4U 0614+09}

\author{Peter Bult}
\affiliation{Astrophysics Science Division, 
  NASA's Goddard Space Flight Center, Greenbelt, MD 20771, USA}
 
\author{Diego Altamirano}
\affiliation{Physics \& Astronomy, University of Southampton, 
  Southampton, Hampshire SO17 1BJ, UK}
 
\author{Zaven Arzoumanian} 
\affiliation{Astrophysics Science Division, 
  NASA's Goddard Space Flight Center, Greenbelt, MD 20771, USA}

\author{Edward M. Cackett}
\affiliation{Department of Physics \& Astronomy, Wayne State University, 
  666 W. Hancock, MI 48201, USA}

\author{Deepto Chakrabarty}
\affil{MIT Kavli Institute for Astrophysics and Space Research, 
  Massachusetts Institute of Technology, Cambridge, MA 02139, USA}

\author{John Doty}
\affil{Noqsi Aerospace Ltd, 2822 S Nova Road, Pine, CO 80470, USA}

\author{Teruaki Enoto}
\affil{Department of Astronomy, Kyoto University, 
  Kitashirakawa-Oiwake-cho, Sakyo-ku, Kyoto, Kyoto 606-8502, Japan}
\affil{The Hakubi Center for Advanced Research, Kyoto University, 
  Yoshida-Ushinomiya-cho, Sakyo-ku, Kyoto, Kyoto 606-8302, Japan}

\author{Keith C. Gendreau} 
\affiliation{Astrophysics Science Division, 
  NASA's Goddard Space Flight Center, Greenbelt, MD 20771, USA}

\author{Sebastien Guillot} 
\affil{CNRS, IRAP, 9 avenue du Colonel Roche, BP
  44346, F-31028 Toulouse Cedex 4, France} 
  ni
\affil{Universit\'e de Toulouse, CNES, UPS-OMP, F-31028 Toulouse, France}

\author{Jeroen Homan}
\affil{Eureka Scientific, Inc., 2452 Delmer Street, Oakland, CA 94602, USA}
\affil{SRON, Netherlands Institute for Space Research,
    Sorbonnelaan 2, 3584 CA Utrecht, The Netherlands}

\author{Gaurava K. Jaisawal}
\affil{National Space Institute, Technical University of Denmark, 
  Elektrovej 327-328, DK-2800 Lyngby, Denmark}

\author{Frederick K. Lamb}
\affil{Center for Theoretical Astrophysics and Department of Physics,
  University of Illinois at Urbana-Champaign,
  1110 West Green Street, Urbana, IL 61801-3080, USA}
\affil{Department of Astronomy, 
  University of Illinois at Urbana-Champaign, 
  1002 West Green Street, Urbana, IL 61801-3074, USA}
  
\author{Renee M. Ludlam}
\affil{Department of Astronomy, University of Michigan, 
  1085 South University Ave, Ann Arbor, MI 48109-1107, USA}

\author{Simin Mahmoodifar} 
\affil{Astrophysics Science Division and Joint Space-Science Institute,
  NASA's Goddard Space Flight Center, Greenbelt, MD 20771, USA}
\affil{CRESST II and the Department of Astronomy, 
  University of Maryland, College Park, MD 20742, USA}
  
\author{Craig Markwardt}
\affiliation{Astrophysics Science Division, 
  NASA's Goddard Space Flight Center, Greenbelt, MD 20771, USA}

\author{Takashi Okajima}
\affiliation{Astrophysics Science Division, 
  NASA's Goddard Space Flight Center, Greenbelt, MD 20771, USA}

\author{Sam Price}
\affiliation{Mission Engineering and Systems Analysis Division, 
  NASA's Goddard Space Flight Center, Greenbelt, MD 20771, USA}

\author{Tod E. Strohmayer} 
\affil{Astrophysics Science Division and Joint Space-Science Institute,
  NASA's Goddard Space Flight Center, Greenbelt, MD 20771, USA}

\author{Luke Winternitz}
\affiliation{Mission Engineering and Systems Analysis Division, 
  NASA's Goddard Space Flight Center, Greenbelt, MD 20771, USA}

\begin{abstract}
  We report on the detection of a kilohertz quasi-periodic oscillation 
  (QPO) with the Neutron Star Interior Composition Explorer 
  (\nicer). Analyzing approximately 165 ks of \nicer exposure on 
  the X-ray burster 4U 0614+09, we detect multiple instances of a 
  single-peak upper kHz QPO, with centroid frequencies that range from 
  400 Hz to 750 Hz. We resolve the kHz QPO as a function of energy, and 
  measure, for the first time, the QPO amplitude below 2 keV. We find 
  the fractional amplitude at 1 keV is on the order of 2\% rms, and 
  discuss the implications for the QPO emission process in the context 
  of Comptonization models. 
\end{abstract}


\section{Introduction} \label{sec:intro}
	Kilohertz quasi-periodic oscillations (QPOs, \citealt{Klis1996,
    Strohmayer1996}) are the fastest variability signatures observed 
    from accreting neutron star X-ray binaries. Ubiquitous across 
    source types, these QPOs may appear as a single or double (twin) 
    peak in the power spectrum, with centroid frequencies that can 
    move from 200 Hz up to 1200 Hz in correlation with rising X-ray
    luminosity \citep{Klis2006}. 
    
    Given the very short timescale variability they represent, it
    is clear that kHz QPOs must originate from close to the neutron 
    star surface, and hence they have drawn much attention. Many of 
    the suggested models for kHz QPOs associate either the lower or 
    upper peak with orbital motion at the inner edge of the accretion disk 
    \citep{Miller1998,Stella1999,Alpar2008,Bachetti2015}.
    Alternative mechanisms have been proposed also
    \citep{Kato2004,Kluzniak2004,Zhang2004}, but so far no
    single model can account for all of the observed properties
    of these oscillations. 
    
    Both observational and theoretical studies of kHz QPOs have 
    focused on their centroid frequencies, and how these change 
    with respect to system parameters such as luminosity or the 
    neutron star spin frequency. Comparatively little attention has been 
    given to how the emergent X-ray flux is being modulated. The 
    reason for this disparity is clear; frequencies provide a direct 
    handle on the dynamics in the accretion system, whereas the 
    modulation mechanism depends also on uncertainties in 
    interpretations of the spectrum.
    
    A path forward is offered by the joint analysis of both 
    spectral and timing characteristics \citep{Gilfanov2003, 
    Avellar2013, Barret2013}. Considering the energy dependence of kHz QPO
    amplitudes and time-lags has the potential to tightly 
    constrain the size and geometry of the modulating medium \citep[e.g.,][]{Kumar2016}, 
    and thereby, indirectly, also the driving dynamical mechanism. 
    Degeneracies in spectral models, however, remain a limiting 
    factor.

	Many spectral models for the emission process of kHz QPOs make
    indistinguishable predictions for the kHz QPO amplitude at
    high photon energies. At low photon energies, however, a specific 
    class of spectral models makes a divergent prediction. If the 
    kHz QPO is generated by 
    a coherent oscillation within some property of the Comptonizing medium,
    such as the temperature or optical
    depth, then the QPO amplitudes should rise at 
    energies below 2 keV \citep{Lee2001}. In contrast, models
    that associate the QPO emission process with luminosity 
    variations, for instance from a boundary layer, predict that
    the QPO amplitude should continue to decrease toward lower 
    energies \citep[see, e.g.,][]{Miller1998}.
    Measurements of kHz QPO properties below 2 keV, however, 
    have so far not been possible due to instrument limitations. 
    
    Launched in 2017 June, the Neutron Star Interior Composition 
    Explorer (\nicer, \citealt{Gendreau2017}) provides good 
    spectral and timing capabilities combined with large 
    collecting area at 1 keV. 
    In this Letter, we present \nicer observations of the neutron 
    star low-mass X-ray binary (LMXB) \osix, a low-luminosity 
    burster \citep{Swank1978} known to show type I X-ray bursts 
    with 415 Hz burst oscillations \citep{Strohmayer2008} and
    kHz QPOs \citep{Ford1997} over a wide range of frequencies 
    \citep{Mendez1997}. Leveraging \nicer's low energy passband,
    we report on the first measurements of kHz QPOs in soft 
    X-rays.
    
\section{Observations} \label{sec:obs}
	The \nicer X-ray Timing Instrument (XTI, \citealt{Gendreau2016})
    consists of 56 co-aligned X-ray concentrator optics, each paired
    with a silicon drift detector sensitive in the $0.2-12$ keV band
    \citep{Prigozhin2012}. The 52 operating detectors collectively 
    provide an effective area of 1900 cm$^2$ at $1.5$ keV, with an energy resolution
    of $\sim100$ eV.
    
    As part of its main science program, \nicer has extensively monitored the
    low-luminosity X-ray burster \osix. In this work, we analyze the data
    collected during the first three months of this campaign, specifically
    between 2017 August 15 and 2017 October 30 (ObsID 1050020101 through 
    1050020131). We processed these data using \textsc{heasoft} 
    version 6.23 and \textsc{nicerdas} version 2018-03-01\_{V003}, 
    using standard filtering criteria: we selected only data collected
    with a pointing offset less than $54\arcsec$, more than $40\arcdeg$
    away from the bright Earth limb, more that $30\arcdeg$ away
    from the dark Earth limb, and outside the South Atlantic 
    Anomaly.
    
    We then constructed a 8-s resolution light curve using the $12-15$ 
    keV energy band. This energy range is higher than the nominal 
    passband of the instrument because the performance of the optics 
    and detectors diminishes such that essentially no astronomical signal 
    is expected above 12 keV. Whenever this light curve had a rate
    greater than 1 ct/s, we observed a correlated increase in the 
    $0.4-12$ keV rate. We therefore attributed those epochs to
    high-background intervals, and removed them from our analysis. About
    $2.5$ ks of high-background exposure was removed in this way.
    After filtering we were left with approximately 165 ks worth of 
    good time exposure.
    
    Because \nicer does not provide imaging capabilities, we determined
    the background contribution from \nicer observations of the blank-field 
    \textit{Rossi X-ray Timing Explorer} (\rxte) background region 8 
    \citep{Jahoda2006}, using the same filtering criteria. Here we note that our source count-rate
    was much higher than the background rate, such that the analysis does not
    depend on our choice of background region. After filtering we obtained
    74 ks of good time exposure for the background field with an averaged 
    background rate of approximately 0.04 counts/s/det in the $0.4-10$ keV band. 

	During these observations, \nicer detected a single type I X-ray burst
    on 7 September 2017 (MJD 58003). We excluded this X-ray burst from
    our analysis. We used a burst epoch starting 50 seconds
    prior to the burst onset, and approximately 375 seconds in length, such
    that it extended to the end of the exposure.

	\begin{figure*}
		\centering
		\includegraphics[width=\linewidth]{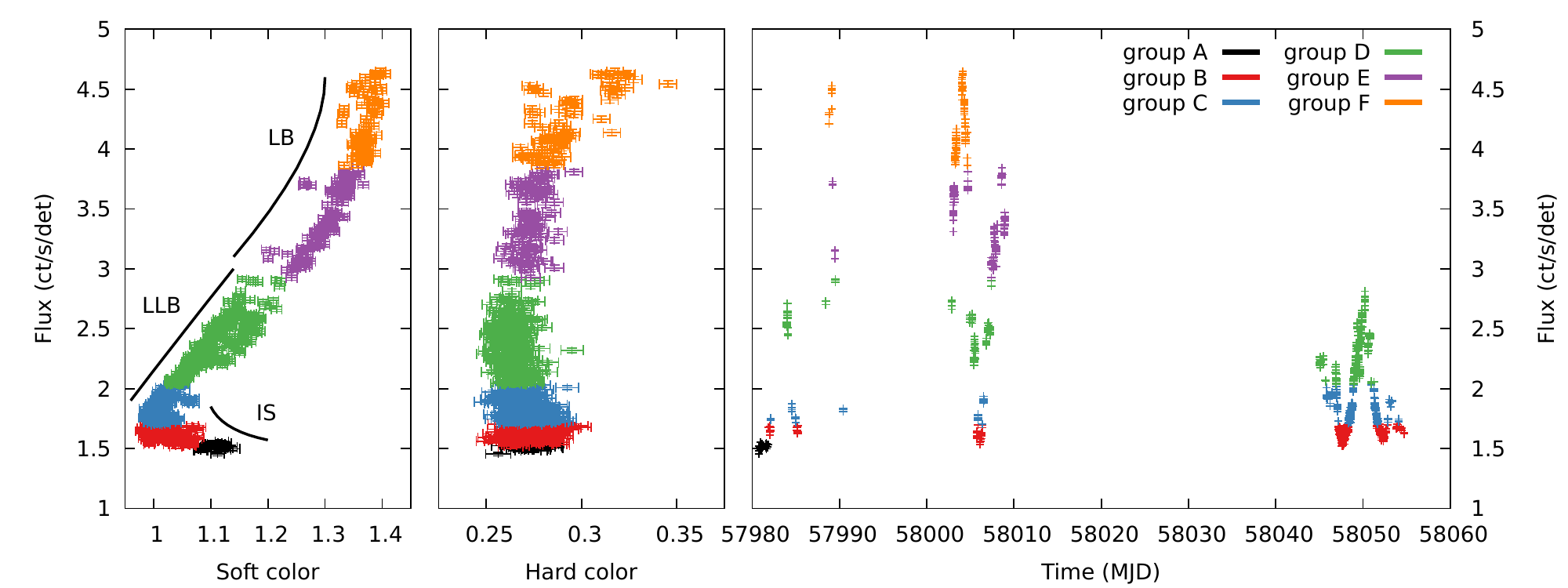}
        \caption{Color evolution of 4U 0614+09, with; left: the
        soft-color intensity diagram; middle: the hard-color intensity
        diagram; and right: the light curve. All panels show the 
        $0.5-6.8$ keV flux. Each point represents a 128-s bin. 
        Color definitions are given in section \ref{sec:colors}.
        See section \ref{sec:timing} for a description of 
        the data grouping and accretion state identification.}
        \label{fig:colors}
	\end{figure*}

\section{Analysis \& Results} \label{sec:results}
	Classified as an `atoll' type source \citep{Hasinger1989},
    the variability properties of \osix have been well established
    by (\rxte; see, e.g., \citealt{Straaten2000, Straaten2002}). As the source luminosity
    changes, it moves through a series of accretion states, each
    with distinct timing and spectral characteristic. The \textit{extreme
    island state} (EIS) shows spectrally hard emission and strong 
    low-frequency variability. At higher luminosity the 
    \textit{island state} (IS) has a softer spectrum and somewhat 
    faster variability. At still higher luminosity the soft emission
    starts to dominate the spectrum and the source moves into the 
    `banana' branch. This branch is sub-divided into three regions: 
    the lower-left `banana' (LLB), where twin kHz QPOs appear; the 
    lower `banana' (LB), which may have only one kHz QPO at a 
    higher frequency; and the upper `banana' (UB), where kHz QPOs 
    are no longer observed. 
    
    The evolution of accretion states is reproduced across many 
    accreting neutron stars \citep{Straaten2005, Altamirano2008b, 
    Bult2015b}, and gives a reliable handle on 
    how to search for kHz QPOs. The accretion state identification 
    for \rxte observations, however, is usually guided by a 
    color ratio centered about 10 keV, which is not readily 
    accessible to \nicer. Hence, we first consider a color 
    analysis that is more appropriate to the \nicer passband.

\subsection{Color analysis} \label{sec:colors}
	We constructed a light curve for the \nicer data using
    a 1/8192-s time resolution. We then divided this light
    curve into 128-s segments, and for each segment, computed
    a soft ($1.1-2.0$ keV / $0.5-1.1$ keV) and hard ($3.8-6.8$ 
    keV / $2.0-3.8$ keV) color ratio, as well as the averaged 
    count-rate ($0.5-6.8$ keV) in that segment\footnote{%
    	We point out that the \nicer \textit{hard}
        color corresponds approximately with the 
        usual \rxte \textit{soft} color.
    }.
    We find that our hard color shows little variation with
    respect to intensity, and provides a poor diagnostic of
    the system's accretion state. Our soft color, on the
    other hand, traces out a pronounced curve as a function 
    of count-rate (Figure \ref{fig:colors}, see section 
    \ref{sec:timing} for data grouping). 

    Over the span of our 74-d light curve the observations of 
    \osix sample approximately ten loops up and down the 
    soft-color-intensity diagram (SID) track, although some
    loops reverse before the highest count-rates are reached. 
    The stable recurrence pattern of this color evolution suggests 
    that a color ratio centered about 1 keV may be a good diagnostic 
    of the accretion state. In this work we therefore use 
    the SID track as our primary indicator for data grouping.

\subsection{Timing analysis}
\label{sec:timing}
    For each 128-s light curve segment in the $2-10$ keV band we computed Fourier 
    transforms, and constructed Leahy-normalized \citep{Leahy1983b} power spectra.
    We used only events above $2$ keV to allow for a clean comparison
    with results obtained using \rxte \citep{Straaten2002}, as
    the low-energy power spectrum can show very different properties
    \citep{Bult2018a}.
    We then grouped segments based on their SID position and
    the shape of their low frequency power spectrum. This results
    in six contiguous data groups along the SID track.
    We designate these groups alphabetically, from the lowest
    to highest count-rates, as `A' through `F'. The SID and 
    light curve for this data grouping are shown in Figure 
    \ref{fig:colors}, along with the hardness-intensity diagram.
    
    For each data group A through F we averaged the individual
    segments to a single power spectrum.
    Inspecting the $2000-4000$ Hz frequency range we find, in each case,
    a mean Leahy power of $2$, as expected for Poisson noise, with no 
    evidence for high frequency signals. Indeed, due to \nicer's modular 
    design, deadtime should not be an issue at 
    the observed count-rates, so we subtract a constant power level of $2$
    from our spectra. Finally, we renormalized the 
    averaged power spectra in terms of fractional rms amplitude with
    respect to the total source count rate \citep{Klis1995}.
    
\begin{figure*}
   		\centering
        \plotone{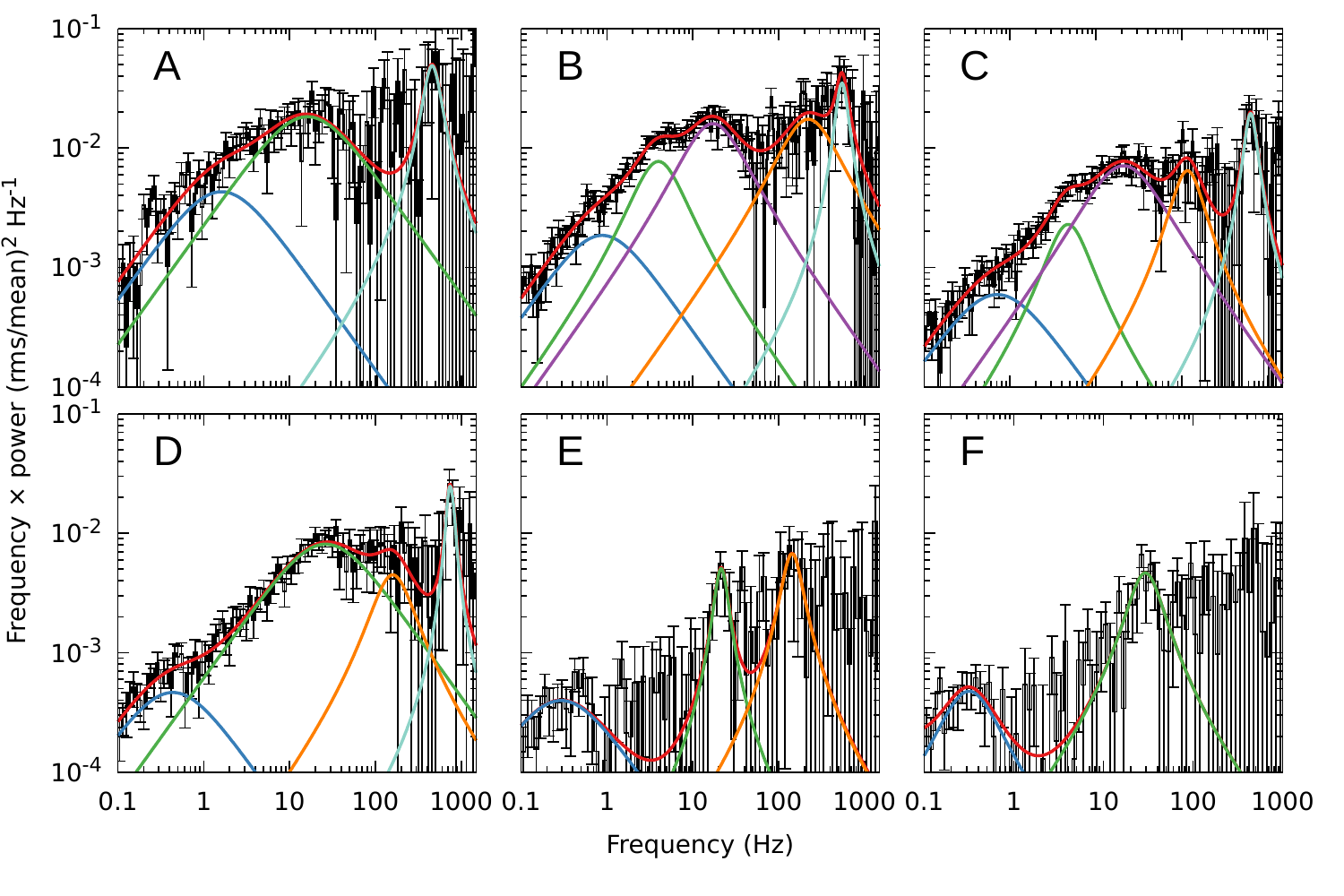}
        \caption{Power spectra of the six data groups as indicated. 
        The red line shows the best fit multi-Lorentzian model, and 
        the remaining colored lines show the model components: break 2 
        (blue), break (green), hump (purple), hHz (orange), 
        and kHz QPO (teal).
        }
        \label{fig:pds}
   	\end{figure*}

    Comparing the averaged \nicer power spectra (Figure \ref{fig:pds}) with 
    results obtained from \rxte \citep{Straaten2002}, we can now
    identify the positions on the SID track with the `atoll' type 
    accretion states.
    At the lowest count-rates (groups A and B) the source populates the 
    island state. Toward higher count-rates (groups C and D) we observe a 
    transition into the lower-left `banana', and finally, at the 
    higher count-rates (groups E and F), we observe `banana' branch
    type variability.
    
    We quantify the power spectra by fitting a multi-Lorentzian model
    \citep{Belloni2002}, with each Lorentzian profile defined as
    $L(\nu ; r, Q, \nu_{\rm max})$, for a characteristic frequency
    $\nu_{\rm max} = \nu_0 \sqrt{ 1 + 1/4Q^2}$, quality factor $Q$, 
    and centroid frequency $\nu_0$. The fractional rms amplitude, 
    $r$, is defined by integrating over the positive frequency 
    domain as
    \begin{equation}
        r^2 = \int_0^{\infty} L(\nu) d\nu.
    \end{equation}
    For all six data groups this model provides an adequate description
    of the power spectrum. Our best-fit results are listed in Table
    \ref{tab:fit}, and the associated power spectra are shown in Figure
    \ref{fig:pds}.
    
\begin{table}[t]
    \newcommand{\mc}[1]{\multicolumn2c{#1}}
    \centering
	\caption{%
    	Power spectrum fit parameters
		\label{tab:fit}	
	}
    \setlength{\tabcolsep}{2pt}
    \begin{tabular*}{\linewidth}{l D D D r }
    \decimals
	\tableline
    ~     & \mc{Frequency} &\mc{Quality}& \mc{Fractional} & \multirow{3}{*}{$\chi^2$ / dof} \\
    ~     & \mc{~}         &\mc{factor} & \mc{amplitude}  &  \\
    ~     & \mc{(Hz)}      &\mc{~}      & \mc{(\% rms)}   &  \\
    \tableline
    \mc{group A | $4$ ks} \\
    \tableline 
    break    &   1.6(0.4)  &  0 ~(fixed) &  11.6(1.4) & \multirow{3}{*}{143/128} \\
    hump     &  16.2(2.8)  &  0 ~(fixed) &  24.0(1.0) & \\
    kHz QPO  & 458. (52)   &  1.9(1.2)   &  19.0(3.5) & \\
    \tableline 
    \mc{group B | $41$ ks} \\
    \tableline
    break   &   0.9(0.3)  &  0.06 (0.14) &   7.5 (1.4) & \multirow{5}{*}{195/180} \\
    break 2 &   4.0(0.3)  &  0.6  (0.2)  &  11.8 (2.0) & \\
    hump    &  17.2(1.1)  &  0.41 (0.14) &  18.5 (1.6) & \\
    hHz     & 222. (49)   &  0.5  (0.3)  &  18.9 (3.1) & \\
    kHz QPO & 548. (24)   &  2.7  (1.3)  &  13.8 (1.7) & \\
    \tableline 
    \mc{group C | $35$ ks} \\
    \tableline
    break 2 &  0.71(0.12)  &  0. ~(fixed)&   4.3(0.3)  & \multirow{5}{*}{137/159} \\
    break   &  4.7(0.5)  &  0.7 (0.3)    &    6.0(1.7) & \\
    hump    & 21.0(1.9)  &  0.27(0.19)   &   13.2(1.5) & \\
    hHz     & 117. (11)   &  1.0 (0.5)   &   8.9(1.5)  & \\
    kHz QPO & 637. (26)   &  2.7 (0.9)   &  10.3(1.1)  & \\
    \tableline 
    \mc{group D | $41$ ks} \\
    \tableline 
    break 2 &   0.44 (0.07)&  0. ~(fixed) &  3.8(0.2) & \multirow{4}{*}{97/117} \\
    break   &  26.3 (2.6)  &  0. ~(fixed) & 15.9(0.7) & \\
    hHz     & 159.  (36)   &  0.8(0.7)    &  8.2(2.2) & \\
    kHz QPO & 748.  (27)   &  4.2(2.0)    &  9.5(1.5) & \\
    \tableline 
    \mc{group E | $15$ ks} \\
    \tableline 
    break 2&   0.29 (0.05)&  0. ~(fixed)&  3.5(0.2) & \multirow{3}{*}{80/95} \\
    break  &  21.5  (1.0) &  2.5 (0.9)  &  5.4(0.6) & \\
    hHz    & 144.   (16)  &  1.7 (1.0)  &  7.5(1.4) & \\
	\tableline
    \mc{group F | $11$ ks} \\
    \tableline
    break 2&  0.32 (0.04) &  0.5(0.2)  &  3.1(0.3) & \multirow{2}{*}{82/95} \\
    break  & 29.3 (2.6)   &  1.0(0.3)  &  7.7(0.6) & \\
    \tableline
	\end{tabular*}
    \flushleft
    \tablecomments{%
     Best fit parameter values for the multi-Lorentzian 
     models that describe the power spectra of the six data 
     groups discussed in Section \ref{sec:timing}. The exposure
     per data group is indicated. Values in parentheses 
     indicate $1\sigma$ uncertainties.  
     }
\end{table}

    We clearly detect kHz QPOs in groups A through D ($>3\sigma$). In
    all cases only a single kHz peak is observed. Based on the
    morphology of the power spectrum, the quality factors of the QPOs,
    and the comparison with \rxte results \citep{Straaten2002,
    Altamirano2008b}, we can identify this single peak as the `upper'
    kHz QPO. The measured frequency correlates with count-rate, and
    moves from $400$ Hz in group A to 750 Hz in group D.
    
    No kHz QPOs are detected in groups E and F. Assuming kHz QPO
    properties similar to those reported for \rxte observations at
    the highest luminosities for this source, that is $Q\simeq10$
    and $\nu_{\rm max}\simeq1000$ Hz \citep{Straaten2002}, we
    obtained a $95\%$ upper limit on a kHz QPO amplitude of $\sim8\%$
    rms in either group.
    
    Although interesting in their own right, a full analysis of the
    broad band power spectrum is beyond the scope of this Letter. 
    Instead we focus our analysis on the energy dependent properties 
    of the kHz QPO.
    
\subsection{Spectral-timing}
    Using the same data grouping described above, we investigated the 
    spectral-timing properties of the kHz QPO by computing 
    time-lags and fractional covariance as a function of energy.
    We calculated these measures by cross-correlating a narrow 
    ($\sim1$ keV) energy band with a broad $0.4-10$ keV reference band, where
    we excluded the band of interest from the reference band 
    \citep{Uttley2014}. For each data group we then integrated
    the cross spectra over frequency intervals the size of the
    QPO full-width-at-half-maximum, $\mbox{fwhm} = \nu_0/Q$,
    centered on the measured kHz QPO centroid frequency. 
    
    \begin{figure}
   		\centering
        \plotone{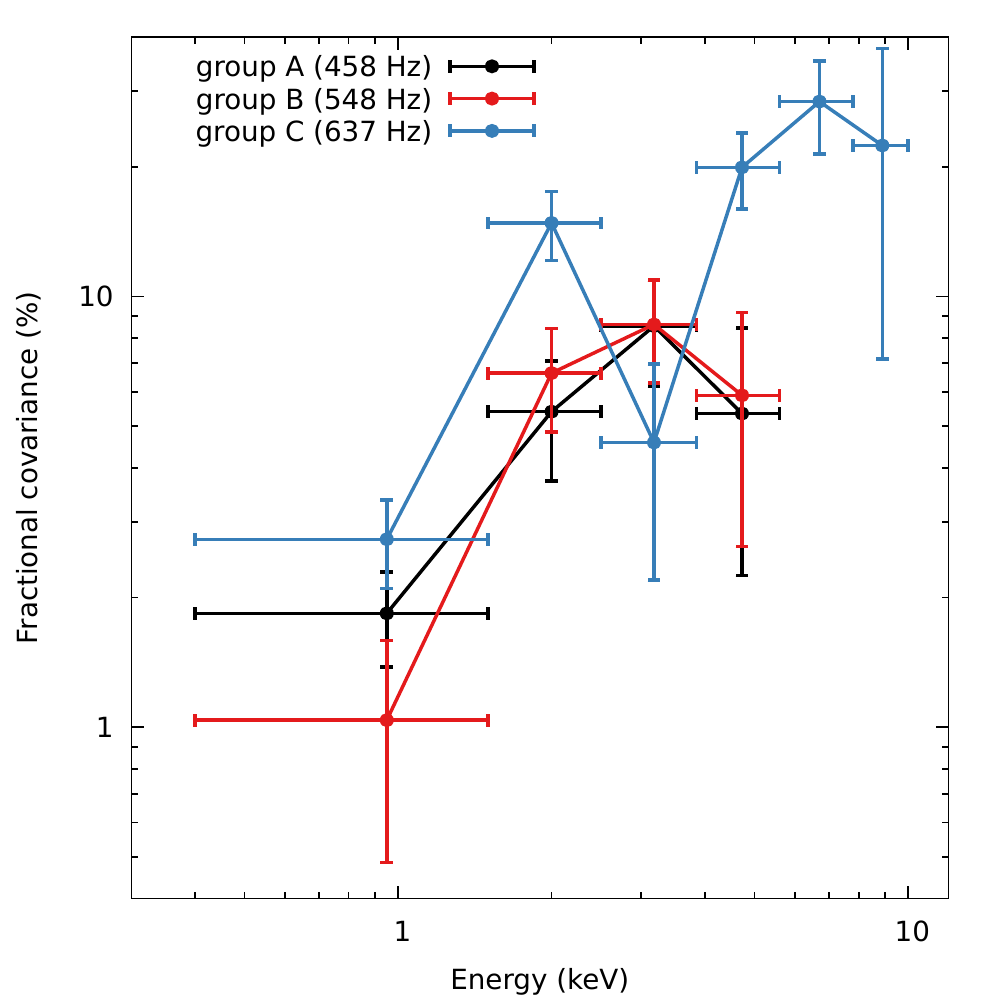}
        \caption{Fractional covariance of the kHz QPO as measured
        in groups A, B, and C. Some of the high energy data points
        for group A and B are not shown as their upper limits are
       	not constraining.}
        \label{fig:cov}
   	\end{figure}
    
    The energy dependent time-lags are statistically consistent with zero,
    with uncertainties on the order of 50 $\mu$s.
    
    The energy dependence of the fractional covariance is shown in Figure \ref{fig:cov}.
    For data groups A through C, we found that the fractional covariance generally
    increases as a function of energy. 
    For data group D, the kHz QPO amplitude was too low to meaningfully 
    constrain the covariance. 
    The increasing trend in amplitude shows some variations, and appears to 
    flatten off, and even turn over above 4 keV.
    We warn, however, that the these changes in amplitude are all smaller than
    the measurement uncertainty. Hence, a larger dataset will be needed to determine
    if such variations are real or merely statistical fluctuations. 
    
    Most notably, we detected the kHz QPO in the lowest energy band, which 
    we defined between 0.4 and 1.5 keV, with a fractional
    covariance of 
    $1.8 \pm 0.5 \%$ rms in group A ($4\sigma$) and at $2.7 \pm 0.6 \%$ 
    rms in group C ($4.3\sigma$). At a $2\sigma$ level, the group B
    measurement is not formally significant.
    We also attempted to perform our measurement at a narrower
    energy resolution by splitting the $0.4-1.5$ keV bin into two
    equal parts. However, the kHz QPO was not significantly
    detected in either sub-bin, with upper limits of $3\%$ rms.

\section{Discussion}
	We analyzed \nicer observations of the low-luminosity burster
    \osix, and detected, for the first time, a kHz QPO at photon
    energies below 2 keV. We argued that this feature corresponds to the
    upper-frequency component of the kHz QPO pair previously identified
    in \rxte observations of accreting neutron stars.
    We found that the kHz QPO amplitude decreases
    rapidly toward low photon energies, with a fractional
    covariance of approximately $2 \%$ rms around 1 keV.
   	Above 2 keV we observed increasing fractional covariance consistent
    with previous findings for the upper kHz QPO \citep{Avellar2013} 
    and with \osix in particular \citep{Troyer2018}. The measured time
    lags are consistent with zero, again in accordance with other
    work \citep{Avellar2013}. 
    
    We only detected the upper kHz QPO, and only in the island state
    and the lower-left portion of the `banana' branch of the atoll-type
    classification. 
    We did not detect kHz QPOs in groups E and F, with an upper
    limit on the QPO amplitudes of $\sim9\%$. Compared to
    the results of \rxte observations, our upper limits are
    similar to reported amplitudes \citep{Straaten2002}, hence
    we cannot rule out the possibility that a narrow kHz QPO
    is present at the highest source luminosities observed with
    \nicer.

    We further note that our data groups cover a large
    timespan, and are potentially susceptible to an observational
    bias.
    In particular, the combination of \nicer's soft passband and diminishing
    QPO amplitudes toward lower energies oblige us to average a 
    large number of observations over week-long intervals to detect 
    the kHz QPO above the noise level. It is well known that, on
    such timescales, secular evolution of the accretion system causes
    the QPO to appear at different frequencies for the same luminosity, 
    giving rise to parallel tracks in flux versus frequency diagrams 
    \citep{Mendez1999}. In averaging many segments, this frequency 
    drift causes the QPO to be smeared out, making it more difficult to detect.  
    This effect is particularly pronounced for narrow QPOs, for
    which the frequency drift may be larger than the QPO
    width. Due to our sampling, this \nicer dataset is therefore 
    more sensitive to the broader kHz QPOs of the island state than 
    it is to the narrow kHz QPOs of the lower `banana' branch.
    
    Our measurements provide a strong challenge to models that associate
    the kHz QPO with a coherent oscillation of the coronal properties,
    such as temperature or optical depth \citep{Lee2001}.
    Any such model predicts that there is a pivot energy, $E_P$, above 
    and below which the QPO amplitude increases. The $E_P$ can be
    anywhere between $1$ and $20$ keV, with the temperature, optical
    depth, and geometry of the Comptonizing medium setting the specific
    value \citep[see, e.g.][]{Miller1992,Lee1998,Kumar2014}.
    Hence, at the lowest energies, the QPO amplitude 
    should have turned over and begun increasing. In contrast to this 
    prediction, we do not observe any such turnover in QPO amplitude.

    In principle it is possible that the spectrum of the kHz QPO
    pivots within the boundaries of our lowest energy bin. 
    In this scenario the QPO phase above and below the pivot energy 
    would be $180\arcdeg$ apart, leading to artificial deconstructive
    interference in our measurement. To test this possibility we
    have split the lowest energy bin of group C in two equal parts, 
    and computed the fractional covariance in each. If the QPO
    indeed pivots, then one would expect either or both of these
    sub-bins to show an increase in amplitude. Contrary to this
    prediction, we do not detect the QPO at a significant level in 
    either sub-bin, with upper limits of $3\%$ fractional covariance. 
    
    A possible way of reconciling the Comptonization model with our 
    measurements comes from allowing a feedback between the Comptonizing
    medium and the thermal component that provides the seed photons
    \citep{Kumar2014, Kumar2016}. If a large fraction of the hard photons
    impinge back on and raise the temperature of the soft photon source,
    then the emergent spectrum no longer shows a low energy pivot. Under
    these conditions, however, the QPO amplitudes would continuously
    increase toward high energies ($>10$ keV), which conflicts with
    observational evidence that shows a flattening of the amplitude
    instead \citep{Berger1996, Peille2015}.
    
    In summary, we have detected the upper kHz QPO below 2 keV,
    and we see no evidence for a low-energy pivot in QPO amplitude. This
    detection poses a challenge to models relying on coherent
    oscillations of coronal properties to explain the radiative
    process responsible for the QPO. Instead, the current body of
    evidence favors an upper kHz QPO associated with an azimuthal
    oscillation, at the inner edge of the disk \citep{Bult2015}, that 
    modulates the luminosity of a boundary layer \citep{Gilfanov2003,
    Peille2015, Troyer2017}.
    
    ~\\

\acknowledgments
This work was supported by NASA through the \nicer mission and the
Astrophysics Explorers Program, and made use of data and software 
provided by the High Energy Astrophysics Science Archive Research Center 
(HEASARC).
P.B. was supported by an NPP fellowship at NASA Goddard Space Flight Center. 
E.M.C. gratefully acknowledges NSF CAREER award AST-1351222.

\facilities{ADS, HEASARC, NICER.}

\bibliographystyle{fancyapj}

\end{document}